\documentclass[final,1p,times]{elsarticle}

\usepackage{amssymb}
\usepackage{booktabs} 
\usepackage{epstopdf}

\journal{Encyclopedia of Big Data Technologies}

\begin{document}

\begin{frontmatter}



\title{Privacy-Preserving Record Linkage}

\author[label1]{Dinusha Vatsalan\corref{cor1}}
\address[label1]{School of Computing, Macquarie University, NSW 2109, Australia}
\address[label3]{School of Science and Technology, Hellenic Open University, Patras, Greece}
\cortext[cor1]{Corresponding author.}
\ead{dinusha.vatsalan@mq.edu.au}

\author[label3]{Dimitrios Karapiperis}
\ead{dkarapiperis@eap.gr}

\author[label3]{Vassilios S. Verykios}
\ead{verykios@eap.gr}

\end{frontmatter}



\noindent
\textbf{Definition}

\noindent
Given several databases containing person-specific data held by different organizations, Privacy-Preserving Record Linkage (PPRL) aims to identify and link records that correspond to the same entity/individual across different databases based on the matching of personal identifying attributes, such as name and address, without revealing the actual values in these attributes due to privacy concerns.

\section{Synonyms}
\label{sec:synonyms}

\noindent
Private Data Matching, Private Record Linkage, Blind Data Linkage, Private Data Integration



\section{Overview}
\label{sec:overview}


In the current era of Big Data personal data about people, such as customers, patients, tax payers, and clients, are dispersed in multiple different sources collected by different organizations. Several applications have begun to leverage tremendous opportunities and insights provided by linked and integrated data. Examples range from healthcare, businesses, social sciences, to government services and national security. Linking data is also used as a pre-processing step in many data mining and analytics projects in order to clean, enrich, and understand data for quality results~\cite{Chr12}.

However, the growing concerns of privacy and confidentiality issues about personal data pose serious constraints to share and exchange such data across organizations for linking. Since a unique entity identifier is not available in different data sources, linking of records from different databases needs to rely on available personal identifying attributes, such as names, dates of birth, and addresses. Known as quasi identifiers (QIDs), these values in combination not only allow uniquely identifying individuals but also reveal private and sensitive information about them~\cite{Vat13}.

Privacy-preserving record linkage (PPRL) aims to identify the same entities from different databases by linking records based on encoded and/or encrypted QIDs, such that no sensitive information of the entities is revealed to any internal (parties involved in the process) 
and external 
(external adversaries and eavesdroppers) 
parties. While a variety of techniques and methodologies have been developed for PPRL over the past two decades, as surveyed in~\cite{Vat13,Vat17b}, this research field is still open to several challenges, especially with the Big Data revolution.


A typical PPRL process involves several steps, starting from pre-processing databases, encoding or encrypting records using the privacy masking functions~\cite{Vat13}, applying blocking or other forms of complexity reduction methods~\cite{Chr12}, then matching records based on their QIDs using similarity functions~\cite{Chr12}, and finally clustering or classifying matching records that correspond to the same entity. 
Additional steps of evaluation of the linkage 
as well as manual review of certain records 
are followed in non-PPRL applications. However, these two steps require more research for PPRL as they generally require access to raw data and ground truth which is not possible in a privacy-preserving context.


Despite several challenges, the output of PPRL could certainly bring in enormous potential in the Big Data era for businesses, government agencies, and research organizations. PPRL has been an active area of research in the recent times and it is now being applied in real-world applications~\cite{Boy15,Ran14}. In this chapter, we will describe some of the key research findings of PPRL and example PPRL applications. We will also discuss directions for future research in PPRL. 

\section{Key research findings}
\label{sec:findings}

\begin{figure*}[t!]
  \centering
  \includegraphics[width=1.0\textwidth]{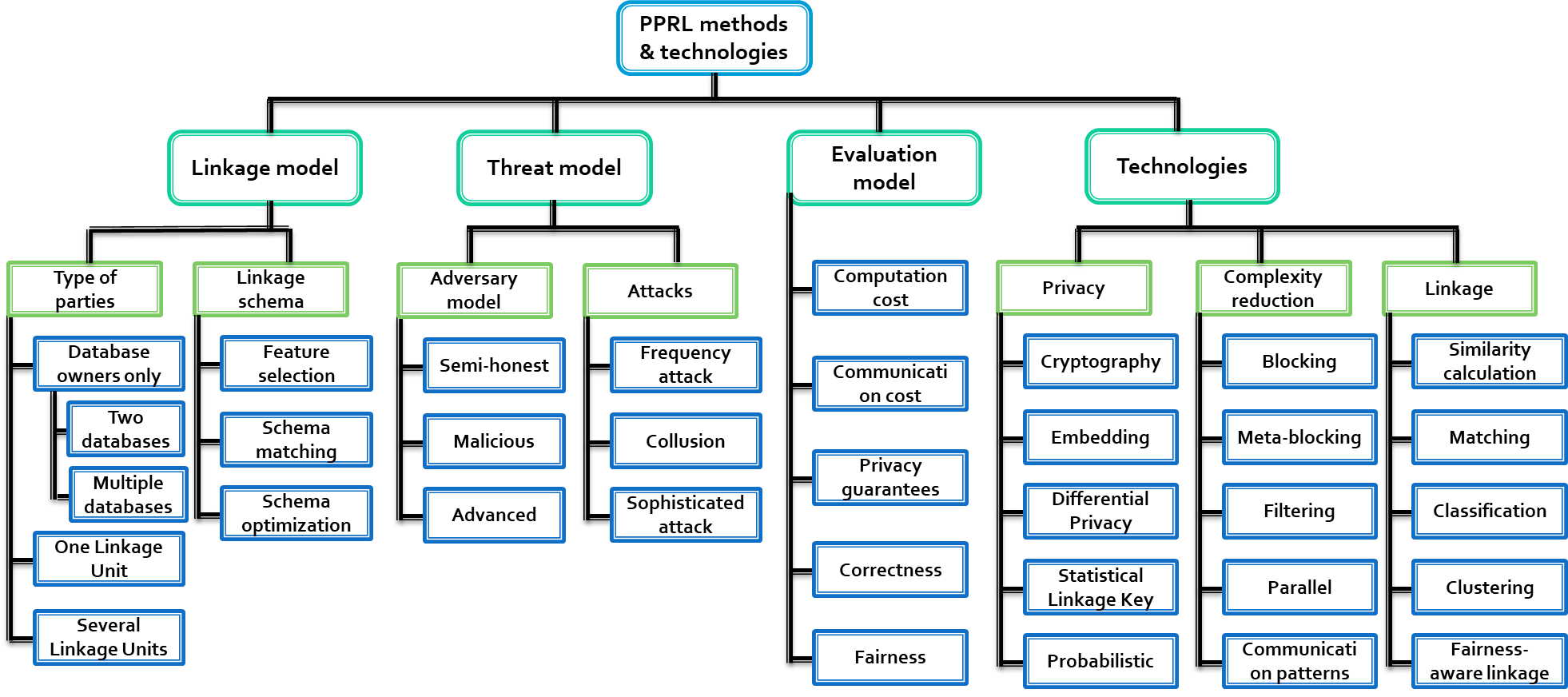}
 
  \caption{A taxonomy of methodologies and technologies used in PPRL.
    }
\label{fig:taxonomy}
\end{figure*}

In this section, we present the key research findings in PPRL with regard to the methodologies and technologies used, as characterized in Figure~\ref{fig:taxonomy}. 
%

\noindent
\subsection{Linkage model}

\noindent
\textbf{Type of parties}:
Considering the number and type of parties involved, the linkage model can be categorized as two-party or multi-party protocols with or without one or several linkage units (LUs).
Two database owners (DOs)/parties involve in two-party protocols in order to identify the matching records in their databases. This linkage model requires more sophisticated techniques to ensure that the parties do not learn any sensitive information about each other's data.
A LU is therefore commonly used in linkage models to conduct/facilitate the linkage. A major drawback of LU-based protocols is that they require a trusted LU in order to avoid possible collusion attacks, where one of the DOs collude with the LU to learn about other DO's data~\cite{Vat13}.
In some models several (more than one) LUs are used, for example, one is responsible for secret keys management, another one for facilitating the complexity reduction step, while matching of records is conducted by a different LU. Separating the tasks across several LUs reduces the amount of information learned by a single party, however collusion could compromise the privacy. 
In multi-party linkage the aim is to identify cluster of records from multiple (more than two) databases that refer to the same entity. 
The linkage process becomes complicated with the increase of number of databases. Processing can be distributed among multiple parties to improve efficiency as well as to improve privacy guarantees by reducing the amount of information learned by a single party.

\noindent
\textbf{Linkage schema}:
The linkage schema dimension consists of feature selection, schema matching, and schema optimization. Different types of QIDs have been used for PPRL, with the most commonly used QIDs include name (string), address (string or text), age (numeric), gender (categorical), and date of birth (date). Schema matching identifies the common schema across different databases~\cite{Sca07}. The success of linkage depends on the parameter setting used which needs to be tuned appropriately in order to optimize the linkage results.
Grid search and random search are two optimization methods, however they tune parameters in an isolated way disregarding past evaluations of parameter combinations~\cite{Ber12}. Bayesian optimization can efficiently bring down the time spent to get the optimal set of parameters~\cite{Sno12} by taking into account the information on the parameter combinations it has previously seen thus far when choosing the parameter set to evaluate next. 

\subsection{Threat model}

\noindent
\textbf{Adversary model}:
Different adversary models are assumed in PPRL methodologies~\cite{Vat13}, which are categorized as honest-but-curious/semi-honest, malicious, and advanced models.
The semi-honest model is the most commonly used adversary model in existing PPRL techniques~\cite{Vat13}, where the parties are assumed to honestly follow the steps of the protocol while being curious to learn from the information they received throughout the protocol. This model is not realistic in real applications due to the weak assumption of privacy against adversarial attacks.
Malicious model, on the other hand, provides a strong assumption of privacy such that the parties involved in the protocol may not follow the steps of the protocol by deviating from the protocol, sending false input, or behaving arbitrarily. However, more complex and advanced privacy techniques are required to make the protocols resistant against such malicious adversaries. Hybrid models, such as accountable computing and covert models, lie in between the semi-honest model, which is not realistic, and the malicious model, which requires computationally expensive techniques~\cite{Vat13}.  

\noindent
\textbf{Attacks}:
Several attacks have been developed for PPRL techniques to investigate the resistance of such techniques to those attacks. Frequency attacks are most commonly used, where the frequency of encoded values are mapped to the frequency of known unencoded values~\cite{Vat14}. Collusion attacks are possible in LU-based and multi-party models where subsets of parties collude to learn another party's data~\cite{Vat14,Ran20}. More sophisticated attacks have been developed against certain privacy techniques. For example, Bloom filters (as described below) are susceptible to cryptanalysis attacks, which allow the iterative mapping of bit patterns back to their original unencoded QID values based on their frequency alignments depending upon the parameter setting used~\cite{Chr17,Kuz11}.

\subsection{Evaluation model}

\noindent
Evaluation of the performance of PPRL consists of five main criteria: \textbf{Computation and communication costs} determine the efficiency aspect that are often measured either theoretically using the big-O notation~\cite{Dan10b} or empirically using runtime, memory size, number of communication steps, number and size of messages to be communicated, and number of comparisons required~\cite{Vat13}. \textbf{Privacy guarantees} are either formally proven or empirically measured using metrics such as Information gain and disclosure risk metrics~\cite{Vat14} against privacy attacks. \textbf{Correctness and fairness} correspond to the linkage quality aspect, where correctness is the accuracy of linkage results measured using precision, recall, area under curve (AUC), and F1-measure~\cite{Chr12}, and fairness is the accuracy of linkage results with regard to different subgroups of individuals~\cite{Zaf15}.

\subsection{Technologies}
\label{sec:tech}

\noindent
\textbf{Privacy technologies}:
We categorize the key privacy technologies as: 

\begin{enumerate}

\item \textbf{Cryptography} refers to secure multi-party computation techniques, such as homomorphic encryptions, secret sharing, and secure vector operations~\cite{Lin09}. These techniques are provably secure and highly accurate, however, they are computationally expensive. 
An example PPRL technique based on cryptographic techniques is the secure edit distance algorithm for matching two strings or genome sequences~\cite{Ata03}, which is quadratic in the length of the strings.


\item \textbf{Embedding techniques} allow data to be mapped into a multi-dimensional metric space while preserving the distances between  original data~\cite{Sca07}. 
It is difficult to determine the appropriate dimensionality for the metric space. 
%
%
A recent work proposed a framework for embedding string and numerical data with theoretical guarantees for rigorous specification of space dimensionality~\cite{Kara17}.

\item \textbf{Differential privacy} is a rigorous definition that 
provides guarantees of indistinguishability of an individual regardless of the presence or absence of the individual's record in the data with high probability. It has been used in PPRL to perturb data by adding noise such that every individual in the dataset is indistinguishable~\cite{Vat14,Vat13}. However, adding noise incurs utility loss and volume increase. Output constrained differential privacy is a recently introduced privacy model for PPRL that allows disclosing matching records while being insensitive to the presence or absence of a single non-matching record~\cite{He17}.

\item \textbf{Statistical linkage key (SLK)} contains derived values from QIDs which is generated using a combination of components of a set of QIDs. 
An example is the SLK-581 consisting of the second and third letters of first name, the second, third and fifth letters of surname, full date of birth, and sex, which was developed by the Australian Institute of Health and Welfare to link records from the Home and Community Care datasets~\cite{Ran16c}. 
A recent study has shown that SLK-based techniques fall short in providing sufficient privacy protection and sensitivity~\cite{Ran16c}. 

\item \textbf{Probabilistic methods} are the most widely used techniques for practical PPRL applications due to their efficiency and controllable privacy-utility trade-off. These methods use probabilistic data structures
for mapping/encoding data such that the actual distances between original data are preserved depending on the false positive probability of the mapping. A recent study has shown that, if effectively used, probabilistic techniques can achieve high linkage quality comparable to using unencoded records~\cite{Ran14}.

For example, Bloom filter encoding is one probabilistic method that has been used 
in several PPRL solutions~\cite{Dur12,Ran14,Sch15,Seh15,Vat12,Vat14c}.
A Bloom filter $b_i$ is a bit array of length $l$ bits
where all bits are initially set to $0$. $k$ independent hash
functions, $h_j$, with $1 \le j \le k$, are
used to map each of the elements $s$ in a set $S$ into the Bloom filter by
setting the bit positions $h_j(s), 1 \le j \le k$ to $1$. 
\begin{figure*}[t!]
  \centering
  \includegraphics[width=1.0\textwidth]{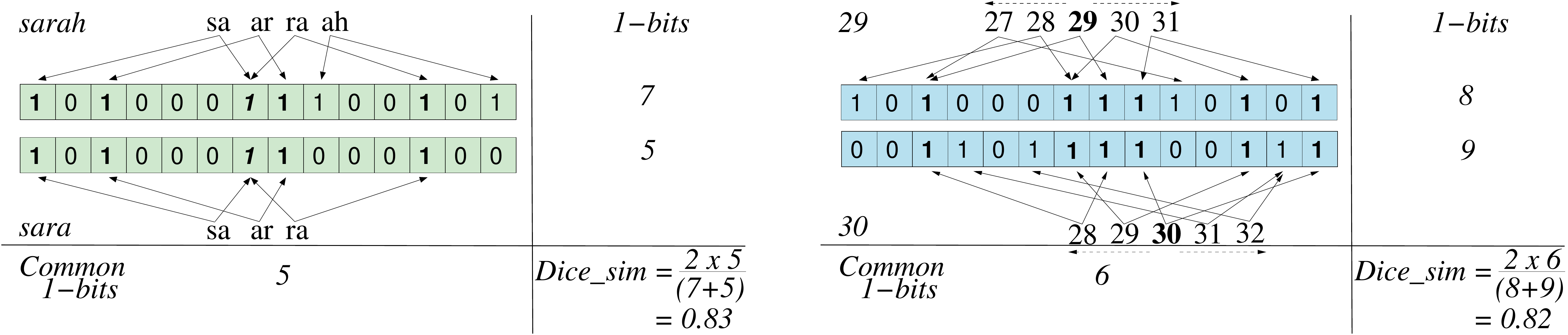}
 
  \caption{Bloom filter-based matching for string~\cite{Sch15} (left) and numerical~\cite{Vat16} (right) data.
    }
\label{fig:BF}
\end{figure*}
%
As shown in Figure~\ref{fig:BF}, the set $S$ of $q$-grams (sub-strings of length $q$) for string QIDs (left) or neighbouring values for numerical QIDs (right) can be hash-mapped into Bloom filters~\cite{Sch15,Vat16}. The resulting Bloom filters can be matched using a similarity function, such as the Dice-coefficient~\cite{Chr12} which is calculated as: 
$Dice\_sim (b_1, \cdots, b_p) = \frac{p \times c}{\sum_{i=1}^{p} x_i}$, 
where $p$ is the number of Bloom filters compared, $c$ is the number of common bit positions that are set to $1$ in
all $p$ Bloom filters, and $x_i$ is the number
of bit positions set to $1$ in $b_i$, $1 \le i \le p$.
The matching can either be done by a LU~\cite{Dur12,Sch15} 
%
or collaboratively by the DOs~\cite{Vat12,Vat14c}.

\end{enumerate}

\noindent
\textbf{Complexity reduction technologies}:
The bottleneck of the PPRL process is the comparison of records across different databases using similarity functions, which is equal to the product of the sizes of the databases. Computational technologies have been used to improve the scalability of PPRL: 

\begin{enumerate}

\item \textbf{Blocking} is defined on selected attributes (blocking keys) 
and it partitions the records in a database into several blocks or clusters based on the blocking key values such that comparison can be restricted to the records of the same block. A variety of blocking techniques has been developed~\cite{Vat13}. Recent examples are randomized blocking methods based on Locality-Sensitive Hashing, 
which provide theoretical guarantees for identifying similar record pairs in the embedding space with high probability~\cite{Dur12,Kar14}.

\item \textbf{Meta-blocking} 
is the process of restructuring a collection of generated blocks to be compared in the next step such that unnecessary comparisons are pruned. Block processing for PPRL only received much attention in the recent years with the aim to improve the scalability in Big Data applications by employing such techniques along with blocking techniques~\cite{Kara15b,Ran16b}.

\item \textbf{Filtering} is an optimization technique for a particular similarity function  
to eliminate pairs/sets of records that cannot meet the similarity threshold for the selected similarity measure~\cite{Seh15,Vat12}. 

\item \textbf{Parallel/distributed processing} for PPRL has only seen limited attention so far. Parallel linkage aims at improving the execution time proportionally to the number of processors~\cite{Dal11,Kar14}. This can be achieved by partitioning the set of all record
pairs to be compared, for example using blocking, and conducting the comparison of the different partitions in parallel on different processors. 

\item \textbf{Advanced communication patterns} can be used to reduce the exponential growth of complexity for multi-party linkage. 
Such communication patterns include sequential, ring by ring, tree-based, and hierarchical patterns. Some of these patterns have been investigated for PPRL on multiple databases~\cite{Vat16b}.

\end{enumerate}

\noindent
\textbf{Linkage technologies}:
A variety of linkage methods and technologies has been used.
\begin{enumerate}
    \item \textbf{Similarity functions} are required for fuzzy/approximate matching of QIDs in order to account for data errors and variations in the QIDs used for linkage. Different similarity functions have been used for different QID data types and different encoding/masking functions. For example, Bloom filter encoding-based PPRL requires token-based similarity functions, such as Jaccard, Hamming, or Dice coefficient functions (as described above)~\cite{Sch15,Dur12}. Similarly, embedding techniques have used Euclidean or edit distance as similarity functions~\cite{Sca07,Kara17}.
    \item \textbf{Matching} techniques determine how the linkage of records needs to be performed. It is a common practice to first de-duplicate records (internally link) within a single database before linking with records from other databases. This is known as one-to-one linking. If the databases are not de-duplicated (i.e.\ they contain multiple records corresponding to the same real-world entity), then many-to-many linking is required. Further, in multi-database linking, subset matching is required in certain applications to identify records that match across any subset of databases (for example, patients visited at least three out of five hospitals)~\cite{Vat20}.
    \item \textbf{Classification} techniques, ranging from simple threshold-based, rule-based, to probabilistic linkage and machine learning, have been used in the PPRL literature~\cite{Vat13}. The aim of these classifiers is to classify the record pairs into `matches' or `non-matches' based on the similarity between their QIDs. While machine learning-based classifiers can provide higher linkage quality, they require training data with ground-truth labels (of `matches' and `non-matches') for supervised techniques.
    \item \textbf{Clustering} is an unsupervised technique that aims to group matching records corresponding to the same real-world entity into one cluster. A recent work studies incremental clustering techniques for multi-party PPRL~\cite{Vat20}.
    \item \textbf{Fairness-aware linkage} is important to ensure fairness in linkage with respect to vulnerable sub-groups of the population. Errors in the linkage will propagate through to the subsequent data analysis. Fairness in the linkage process enables fair upstream data analysis. Fairness-aware classification and clustering algorithms have been developed in the literature~\cite{Meh19}, however, mitigating fairness-bias specifically in PPRL has not yet been studied.
\end{enumerate}

\section{Examples of application}
\label{sec:application}

Linking data is increasingly being required in a number of application areas~\cite{Chr12}. 
When databases are linked within a single organization,  
then generally privacy and confidentiality are not of great concern. 
However, linking data from several organizations imposes legal and ethical constraints on using personal data for the linkage, as described by the Australian Data-Matching Program Act\footnote{https://www.legislation.gov.au/Details/C2016C00755}, the EU General Data Protection Regulation\footnote{https://gdpr.eu/}, and the HIPAA Act in the USA\footnote{http://www.hhs.gov/ocr/privacy/}.
PPRL is therefore required in several real applications, as the following examples illustrate:

\begin{enumerate}

\item{\textbf{Healthcare applications}}: 
Several healthcare applications ranging from health surveillance, epidemiological studies, and clinical trials, to public health research require PPRL as an efficient building block. 
For example, a study on surgical treatment received by aboriginal and non-aboriginal people with lung cancer linked data from hospitals and clinical registries with data from central cancer registries and from the Australian Bureau of Statistics using PPRL techniques~\cite{Con04}. Bloom filter-based PPRL was used to link data from several cantonal and national registries in Switzerland to investigate long-term consequences of childhood cancer~\cite{Kue11}. 
In 2016, the Interdisciplinary Committee of the International Rare Diseases Research Consortium launched a task team to explore approaches to PPRL for linking several genomic and clinical data sets~\cite{Bak18}. 


\item{\textbf{Government services and research}}: 
The traditionally used small-scale survey studies have been replaced by linking databases to develop policies in a more efficient and effective way~\cite{Kel02}. The research program `Beyond 2011' established by the Office for National Statistics in the UK, for example aimed to study the options for production of population 
statistics for England and Wales, by linking anonymous data~\cite{ONS13}. 
Social scientists use PPRL in the field of population informatics to study insights into our society~\cite{Kum13b}. 


\item{\textbf{Business collaboration}}:
Many businesses take advantage of linking data across different organizations, such as suppliers, retailers, wholesalers, and advertisers, for improving efficiency, targeted marketing, and reducing costs of their supply chains. 
PPRL can be used for cross-organizational collaborative decision making which involves a great deal of private information that businesses are often reluctant to disclose~\cite{Zhu17}.

\item{\textbf{National security applications}}:
PPRL techniques are being used by national security agencies and crime investigators to identify individuals who have committed fraud or crimes~\cite{Phu12}. These applications integrate data from various sources, such as law enforcement agencies, Internet service providers, the police, tax agencies, government services, as well as financial institutions to enable the accurate identification of crime and fraud, or of terrorism suspects.

\end{enumerate}

\section{Future directions for research}
\label{sec:future_research}

\begin{figure}[t!]
  \centering
  \includegraphics[width=0.43\textwidth]{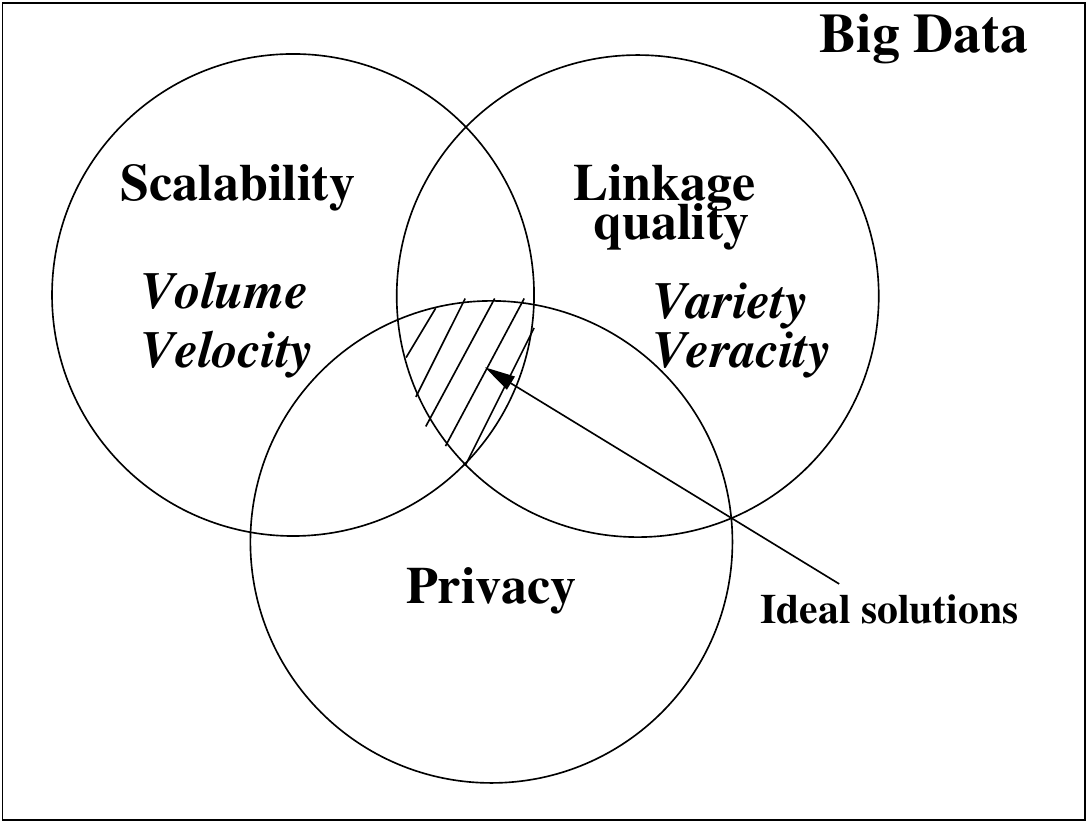}
 
  \caption{Challenges of PPRL for Big Data.
    }
\label{fig:challenges}
\end{figure}

In this section we describe the various open challenges and research directions of PPRL for Big Data applications, as categorized in Figure~\ref{fig:challenges}.


\subsection{Scalability}
\label{sec:chall_scalability}

The \emph{volume} of data increases both due to the large size of databases and the number of different databases. 
Challenges of linking large databases have been addressed by using computational techniques to reduce the number of required comparisons between records. 
However, these techniques do not solve the scalability problem for Big Data completely. For example, even small blocks of records resulting from a blocking technique can still lead to a large number of comparisons between records with the increasing volume of data. Moreover, only limited work has been done to address the challenges of linking multiple databases. 

Scalable blocking techniques are required that can generate blocks of suitable sizes across multiple databases, as are advanced filtering techniques that effectively prune potential non-matches even further. 
With multiple databases, identifying subsets of records that match across only a subset of databases (for example, patients in three out of five hospital databases) is even more challenging due to the large number of combinations of subsets. 
Advanced communication patterns, distributed computing, and adaptive comparison techniques are required towards this direction for scalable PPRL applications. These techniques are largely orthogonal so that they can be combined to achieve maximal efficiency.

Another major aspect of Big Data is the \emph{velocity}, i.e. the dynamic nature of data. Current techniques are only applicable in batch-mode on static and well defined relational data. Required are approaches for dynamic data, real-time integration, and data streams, for adaptive systems to link data as they arrive at an organization, ideally in (near) real-time.
     

\subsection{Linkage Quality}
\label{sec:chall_classify}

Big Data are often complex, noisy, and erroneous, which refer to the \emph{veracity} and \emph{variety} aspects. The linkage can be affected by the quality of data, especially in a privacy-preserving context where linkage is done on the masked or encoded version of the data. Masking data reduces the quality of data to account for privacy and hence the quality of data would have an adverse effect on the whole PPRL process leading to low linkage quality. The trade-off between quality and privacy needs to be handled carefully for different privacy masking functions. 
The affect of poor data quality on the linkage quality becomes worse with increasing number of databases. 
More advanced classification techniques, such as collective~\cite{Bha07} and graph-based~\cite{Kal06} techniques, need to be investigated to address the linkage quality problem of PPRL. 

Developing classifiers that are fair with respect to a protected/sensitive feature~\cite{Zaf15}, such as gender or race, is an important problem for classification in general and specifically for record linkage. Fairness of a classifier with regard to a certain protected feature determines how much the classifier distorts from producing correct predictions with equal probabilities for individuals across different protected groups/values. There has been increased interest in this field due to the concerns that classifiers may introduce significant bias towards certain minority or vulnerable group with regard to the protected feature, such as race or gender, for example against black people in fraud and crime detection systems~\cite{Flo16,Lar16} or against women in job recommendation systems~\cite{Dat15}. However, fairness has not been studied specifically for PPRL so far.


Assessing the linkage quality in a PPRL project is very challenging because it is generally not possible to inspect linked records due to privacy concerns. Knowing the quality of linkage is crucial in many Big Data applications such as in the health or security domains. An initial work has been done on interactive PPRL~\cite{Kum13} where parts of sensitive values are iteratively revealed for manual assessment in such a way that the privacy compromise is limited. Implementing such approaches in real applications
is an open challenge that must be solved. Using heuristic measures to approximately evaluate the linkage quality is another option that requires more research.


%


\subsection{Privacy}
\label{sec:chall_security}

Another open challenge in PPRL is how resistant the techniques are against different adversarial attacks. Most work in PPRL assume the \emph{semi-honest} adversary model~\cite{Lin09} and the trusted LU-based model. Furthermore, these works assume that the parties do not collude with each other~\cite{Vat13}. 
Only few PPRL techniques consider the \emph{malicious} adversary model as it imposes a high complexity~\cite{Vat13}. 
More research is therefore required to develop novel security models that lie between these two models and prevent against collusion risks for PPRL.
%
%


Sophisticated attack methods~\cite{Chr17,Kuz11} have been recently developed that exploit the information revealed during the PPRL protocols to iteratively gather information about sensitive values. 
%
Therefore, existing PPRL techniques need to be hardened to ensure they are not vulnerable to such attacks~\cite{Sch15}. 
Evaluation of PPRL techniques is challenged due to the absence of benchmarks, datasets, and frameworks. Different measurements
have been used~\cite{Dur12,Vat14,Vat13}, making the
comparison of different PPRL techniques difficult.
Synthetic datasets generated with  real data characteristics using data generators~\cite{Tra13} have been used as an alternative to benchmark datasets. This limits the evaluation of PPRL techniques to assess their application in the real setting. 
%

Further, there has not been much interaction between practitioners and researchers of PPRL to allow better understanding of the whole data life cycle and to evaluate the applicability of PPRL in real applications. A comprehensive privacy strategy is essential including a closely aligned PPRL and privacy-preserving data technologies for strategic Big Data applications.






%
%
%

\bibliographystyle{elsarticle-harv}
\bibliography{paper} 

\end{document}